&pdflatex
\documentclass[12pt]{article}
\usepackage{amsmath}
\usepackage{graphicx,psfrag,epsf}
\usepackage{enumerate}
\usepackage{natbib}
\usepackage{url} 
\usepackage{color}

\newcommand{\blind}{0}

\begin{document}

\bibliographystyle{natbib}

\def\spacingset#1{\renewcommand{\baselinestretch}%
{#1}\small\normalsize} \spacingset{1}


\if0\blind
{
  \title{\bf Large Scale Replication Projects in Contemporary Psychological Research}
  \author{{\em Forthcoming in } The American Statistician \\
  \phantom{a} \\
  Blakeley B. McShane\thanks{Correspondence concerning this article should be addressed to Blakeley B. McShane, Marketing Department, Kellogg School of Management, Northwestern University, 2211 Campus Drive, Evanston, IL 60208. E-mail: b-mcshane@kellogg.northwestern.edu. We thank the Defense Advanced Research Projects Administration (agreement number D17AC00001) for partial support of Andrew Gelman's work.}\hspace{.2cm}, Jennifer L. Tackett, and Ulf B\"ockenholt \\
    Northwestern University\\
    and \\
    Andrew Gelman \\
    Columbia University}
  \maketitle
} \fi

\if1\blind
{
  \bigskip
  \bigskip
  \bigskip
  \begin{center}
    {\LARGE\bf Large Scale Replication Projects in Contemporary Psychological Research}
\end{center}
  \medskip
} \fi

\bigskip
\begin{abstract}
Replication is complicated in psychological research because studies of a given psychological phenomenon can never be direct or exact replications of one another, and thus effect sizes vary from one study of the phenomenon to the next---an issue of clear importance for replication. Current large scale replication projects represent an important step forward for assessing replicability, but provide only limited information because they have thus far been designed in a manner such that heterogeneity either cannot be assessed or is intended to be eliminated. Consequently, the nontrivial degree of heterogeneity found in these projects represents a lower bound on heterogeneity.

We recommend enriching large scale replication projects going forward by embracing heterogeneity. We argue this is key for assessing replicability: if effect sizes are sufficiently heterogeneous---even if the sign of the effect is consistent---the phenomenon in question does not seem particularly replicable and the theory underlying it seems poorly constructed and in need of enrichment. Uncovering why and revising theory in light of it will lead to improved theory that explains heterogeneity and increases replicability. Given this, large scale replication projects can play an important role not only in assessing replicability but also in advancing theory.
\end{abstract}

\noindent%
{\it Keywords:} replication, psychology, $p$-value, null hypothesis significance testing, hierarchical, multilevel, meta-analysis, heterogeneity, between-study variation
\vfill

\newpage
\spacingset{1.45} 

\section{Introduction}

The validity of research in the biomedical and social sciences is under intense scrutiny at present, with published findings failing to replicate at an alarming rate. This problem appears particularly acute in psychology, where the failure to replicate several prominent findings (for example, \cite{BarCheBur96,Bargetal01,CarCudYap10,Bem11}) has attracted the attention of both academics and the popular press.

One positive development that has emerged from this unfortunate situation is an increased interest in planning and conducting replications, as evidenced by {\em inter alia} large scale replication projects such as the Many Labs project \cite[]{ManyLabs14}, the Open Science Collaboration (OSC) project \cite[]{OSC2015}, and Registered Replication Reports (RRRs; \cite{SimHolSpe14}) in which one or more psychological phenomena is investigated across multiple laboratories. The projects involve heroic coordination efforts and require tremendous resources but offer the promise of allowing for the assessment of the replicability of individual phenomena and, perhaps taken together, psychological research as a whole. 

Nonetheless, replication is complicated in psychological research because studies of a given psychological phenomenon can never be direct or exact replications of one another \cite[]{Ros91,TsaKwa99,Branetal14,StroStr14,FabWeg16}. Instead, studies differ at minimum in their method factors, that is, anything---known or unknown---pertaining to the implementation of the study that is not directly related to the theory under investigation. Method factors can include seemingly major factors such as the operationalization of the dependent measure(s), the operationalization of the experimental manipulation(s), and unaccounted for moderators but also seemingly minor factors such as the social context, the subject pool, and the time of day (for a comprehensive list, see \cite{Broetal14}). Differences in method factors result in heterogeneity, that is, effect sizes that vary from one study of a given phenomenon to the next---an issue of clear importance for replication.

While current large scale replication projects represent an important step forward for assessing replicability, they provide only limited information. Specifically, and consistent with the fact that heterogeneity has heretofore been underappreciated in psychology, these projects have thus far been designed in a manner such that heterogeneity either cannot be assessed or is intended to be eliminated. Consequently, the nontrivial degree of heterogeneity found in these projects represents a lower bound on heterogeneity. Given this, we recommend enriching large scale replication projects going forward by embracing heterogeneity through the systematic variation of method factors. We argue this is key for assessing replicability: if effect sizes are sufficiently heterogeneous from one study to the next---even if, for example, the sign of the effect is consistent---the phenomenon in question does not seem particularly replicable and the theory underlying it seems poorly constructed and in need of enrichment. In particular, what are believed to be method factors either interact with theoretical moderators or are in fact unaccounted for theoretical moderators. Uncovering this and revising theory in light of it will lead to improved theory that explains heterogeneity and increases replicability. Given this, large scale replication projects can play an important role not only in assessing replicability but also in advancing theory---a fact which has been overlooked to date.

In the remainder of this paper, we briefly review current large scale replication projects. We then discuss assessments of heterogeneity based on them, offer recommendations for enriching them, and discuss challenges with large scale replication in other domains of psychological research. We conclude with a brief discussion.

\section{Large Scale Replication Projects}

\subsection{Many Phenomena, One Study: The Open Science Collaboration Project}

The OSC project is a large scale replication project consisting of replications of 100 social and cognitive psychology phenomena published in three journals ({\em Psychological Science}, {\em Journal of Personality and Social Psychology}, and {\em Journal of Experimental Psychology: Learning, Memory, and Cognition}) and conducted by 270 authors. Fundamental to the approach was that each phenomenon was replicated only once thus allowing for broad coverage across a host of phenomena.

The OSC authors examined five indicators of replicability, and results were, broadly speaking, consistent across them:
\begin{quote}
Replication effects were half the magnitude of original effects, representing a substantial decline. Ninety-seven percent of original studies had statistically significant results. Thirty-six percent of replications had statistically significant results; 47\% of original effect sizes were in the 95\% confidence interval of the replication effect size; 39\% of effects were subjectively rated to have replicated the original result; and if no bias in original results is assumed, combining original and replication results left 68\% with statistically significant effects.
\end{quote}
In sum, the results seem grim.

The major strength of the OSC project is its broad coverage of studies across social and cognitive psychology. However, this comes with a major limitation: because each phenomenon was replicated only once, the OSC project replication setting is the canonical and classical one of a single ``original'' study and a single ``replication'' study of each phenomenon.

Because there is only a single replication study of each phenomenon, heterogeneity cannot be assessed and accounted for. While this is a limitation in and of itself, it also means many of the indicators of replicability discussed above necessarily ignore heterogeneity and are thus miscalibrated.

For example, consider the second---that only 36\% of replication studies had statistically significant results whereas 97\% of the original studies did---which has gained traction in the popular press and reflects the classic definition of replication employed in practice (i.e., a subsequent study is considered to successfully replicate a prior study if either both fail to attain statistical significance or both attain statistical significance and are directionally consistent). When heterogeneity exists but is ignored, the Type I error of the single study significance tests on which these 36\% and 97\% figures are based is inflated above the nominal size $\alpha$ and standard sample size and power formulae overstate power and thus understate Type II error \cite[]{McSBoc17}.

\subsection{One Phenomenon, Many Studies: The Many Labs Project and Registered Replication Reports}

Overcoming the major limitation of the OSC project are the Many Labs project and RRRs which feature multiple replication studies of one phenomenon (RRRs) or a small number of phenomena (Many Labs) across multiple independent laboratories thus allowing for deeper examination of these phenomena. In particular, the Many Labs project investigates sixteen classic and contemporary psychological research phenomena across thirty-six independent samples totaling 6,344 subjects; each of the thirty-six laboratories involved in the project used identical materials and administered them through a web browser in order to ensure procedural consistency across laboratories. Similarly, RRRs consist of ``a set of studies from a variety of laboratories that all followed an identical, vetted protocol designed to reproduce the original method and finding as closely as possible'' \cite[]{SimHolSpe14} and have thus far investigated phenomena that include the facial feedback hypothesis, ego depletion, the effect of time pressure on cooperative decisions, and the link between commitment to and betrayal of a romantic relationship \cite[]{RRRwag16,RRRhagg16,RRRbouw17,RRRcheu16}. Thus, the Many Labs and RRR approach allows data to be integrated via meta-analysis---with the data from each laboratory treated as an independent replication study---to provide more definitive and informative inferences and conclusions.

The primary indicator of replicability used by the Many Labs and RRR authors is whether a meta-analysis that pools the data across the multiple replication studies of a given phenomenon matches the statistical significance and direction of the original study. Empirically, evaluations of replicability diverge between the Many Labs and RRR authors, with the Many Labs authors replicating fourteen of the sixteen phenomena they examined and RRR authors generally failing to replicate phenomena thus far. Rather than reflecting any differences in, for example, the replication approach taken, we believe this divergence more likely reflects the choice of phenomena examined by the Many Labs and RRR authors. 

The major strength of the Many Labs and RRR approach is that pooling data across multiple replication studies allows for more powerful evaluations of replicability---ones that allow heterogeneity to, at least to some degree, be assessed and accounted for. However, this comes with a major limitation: due to the vast resources required to conduct replications across multiple laboratories, the approach is necessarily more focused on one (or a small number of) phenomena and thus does not provide broad coverage across psychology or domains of it. In sum, the strengths and limitations of the Many Labs and RRR approach are opposite those of the OSC project. Yet, like the OSC project, this approach has thus far defined replication in terms that focus on the sign and statistical significance of results.

\section{Assessing Heterogeneity in Large Scale Replication Projects}

In this section, we discuss assessments of heterogeneity---that is, variation in effect sizes from one study of a given phenomenon to the next resulting from differences in method factors---based on large scale replication projects, in particular those following the one phenomenon, many studies approach (i.e., because those following the many phenomena, one study approach do not allow for such assessments). While by no means the only or necessarily the best measure, we here discuss heterogeneity in terms of the $I^2$ statistic (i.e., the proportion of the variation in observed effect sizes due to heterogeneity rather than sampling error) because it is easily interpretable and comparable across phenomena. To place $I^2$ in context, \cite{Pig12} defines low, medium, and high heterogeneity in psychological research as $I^2$ of 25\%, 50\%, and 75\% respectively (see also \cite{HigTho02}). We use this relative measure of heterogeneity here because it facilitates comparison across different phenomena. However, as we discuss in the next section, the absolute level of heterogeneity is also relevant for assessing replicability.

Heterogeneity has generally been regarded as important across sets of studies that consist of general (i.e., systematic or conceptual) replications, and a host of recent data convincingly demonstrates that it is indeed rife and large across them. Consider, for example, the sets of studies included in the comprehensive meta-analyses of the sort published in {\em Psychological Bulletin}, the premier outlet for meta-analyses in psychology. \cite{vanEetal17} examined heterogeneity estimates from 705 meta-analyses published there between 1990 and 2013 and found a median $I^2$ of 71\%. Similarly, \cite{StaCarDou17} surveyed 200 recent meta-analyses published there and found a median $I^2$ of 74\%.

While the finding that heterogeneity is so large in these meta-analyses may perhaps not be so surprising given that the studies included in them tend to feature tremendous variation in their method factors, more surprising---as well as more interesting and more important---is that heterogeneity can also be substantial even across sets of studies that consist entirely of close replications (i.e., studies that use identical or very similar materials). In particular, heterogeneity persists, and to a reasonable degree, even in large scale replication projects such as the Many Labs project and RRRs where rigid, vetted protocols with identical study materials are followed across laboratories in a deliberate attempt to eliminate differences in method factors and thus heterogeneity.

Specifically, random effects meta-analyses conducted by the Many Labs authors yielded nonzero estimates of heterogeneity for all fourteen of the phenomena they found to be non-null and the average $I^2$ across these phenomena was 40\% (see Table 3 of \cite{ManyLabs14}). In addition, among the 6,344 Many Labs subjects were 1,000 recruited via Amazon's Mechanical Turk. The study materials were administered to these 1,000 subjects over seven unique days, beginning on August 29, 2013, and ending on September 11, 2013 (i.e., seven consecutive days excluding Fridays, weekends, and the Labor Day holiday). Restricting attention to only these subjects and treating each unique day as a separate sample yields seven extremely close replications of each phenomenon. Again, however, despite the extreme degree of closeness, heterogeneity is nontrivial: random effects meta-analyses yield nonzero estimates of heterogeneity for nine of the fourteen non-null phenomena, and across these, the average $I^2$ was 21\% \cite[]{McSBocHan16}. Similarly, a moderate amount of heterogeneity has been found in RRRs (see, for example, \cite{RRReerl16} and \cite{RRRhagg16}). In sum, despite protocols explicitly designed to eliminate differences in method factors and thus heterogeneity, heterogeneity was nonetheless nontrivial in large scale replication projects following the one phenomenon, many studies approach.

This is astounding not only from a substantive perspective but also from a purely statistical perspective. In particular, when heterogeneity is in fact nonzero but there are a relatively small number of studies / laboratories (as is unfortunately the case in the Many Labs project and RRRs), standard estimators of heterogeneity are biased downwards and estimates of zero heterogeneity result implausibly often (see, for example, \cite{Chuetal13} and \cite{ChuRabCho13}). For this and related reasons, the Type I error of the significance test of zero heterogeneity is often inflated above the nominal size $\alpha$ and power is low \cite[]{Heudetal06,IoaPatEva07}.

Based on this poorly calibrated significance test, many have wrongly concluded that large scale replication projects do not provide evidence for heterogeneity. For instance, the Many Labs authors write ``tests of heterogeneity suggested that most of the variation in effects is attributable to measurement error'' (by which we believe they mean sampling error as their analyses do not account for measurement error). In contrast, we note that, even if one believes that homogeneity is plausible and that significance testing is a reasonable thing to do, one should be very skeptical of this particular significance test unless the number of studies is indeed very large (i.e., to allow for sufficient power). Further, laying aside any skepticism of significance testing, we note that the privileged role given to the null hypothesis of homogeneity by this particular significance test does not seem reasonable in psychological research because studies of a given psychological phenomenon can never be direct or exact replications of one another; instead, rather than assuming homogeneity and only rejecting it given sufficient evidence to the contrary, it seems much more reasonable to assume some degree of heterogeneity and only conclude in favor of homogeneity given sufficient evidence for it.

Given that large scale replication projects have typically shared their data even at the most granular level, there is great opportunity to move beyond poorly-performing estimators and significance tests in order to better assess heterogeneity. For example, consider again the Many Labs project. Current assessments of heterogeneity are based on the standard two-parameter (one intercept, one variance component) so-called random effects meta-analytic model fit to a single summary statistic computed from the individual-level data from each of the thirty-six laboratories and fit to each of the sixteen phenomena examined by the project entirely separately. Instead, it would be far superior to base assessments of heterogeneity on a hierarchical (or multilevel) model fit to the individual-level observations across all laboratories and phenomena jointly. The partial pooling, particularly of the variance components, possible with such a model would yield much more reliable and precise estimates.

In sum, in large scale replication projects such as Many Labs and RRRs, we should---for substantive reasons (i.e., protocols designed to eliminate heterogeneity) and statistical reasons (i.e., estimators and significance tests that perform poorly in a manner that falsely suggests homogeneity)---expect to observe little to no heterogeneity. The very fact we observe a nontrivial degree of it is compelling evidence that heterogeneity is not only the norm but also cannot be avoided in psychological research---even if every effort is taken to eliminate it.

\section{Enriching Large Scale Replication Projects}

In this section, we make four recommendations pertaining to the design of large scale replication projects, the statistical analysis of the data resulting from them, and assessments of the replicability of the phenomena investigated by them. We then provide an illustration based on the anchoring effect \cite[]{JacKah95} of how these recommendations can improve theory in a manner that explains heterogeneity and increases replicability.

Because heterogeneity cannot be avoided in psychological research and because assessing and accounting for heterogeneity is not possible when analyzing single studies in isolation but only when analyzing multiple studies jointly, our {\em first recommendation} is that future large scale replication projects follow the one phenomenon, many studies approach of the Many Labs project and RRRs rather than the many phenomena, one study approach of the OSC project. This offers substantive as well as statistical benefits: beyond allowing for the ``mere'' assessment of heterogeneity, it can suggest that method factors either interact with theoretical moderators or are in fact unaccounted for theoretical moderators (as discussed in greater depth below); it also provides better calibration of Type I and Type II error and yields more efficient estimates of overall average effects \cite[]{McSBoc14,McSBoc17}.

Before proceeding to our second recommendation, it is important to understand that the degree of heterogeneity across set of studies depends critically on the variation in the method factors across the set. For example, it is unsurprising that studies from the Many Labs project and RRRs are less heterogeneous than studies included in meta-analyses published in {\em Psychological Bulletin} as the former used identical materials while the latter did not. Similarly, it is unsurprising that the Mechanical Turk subsample of the Many Labs sample is less heterogeneous than the full sample as, for example, differences in the subject population across days on Mechanical Turk are likely to be smaller than those across the thirty-six Many Labs. Put differently, the Mechanical Turk subsample features less variation in method factors than the full Many Labs sample which in turn features less variation in method factors than the published meta-analyses. In sum, the less (more) varied a set of studies are in their method factors, the smaller (larger) heterogeneity will be.

Given this, the Many Labs and RRR approach---due to the rigidness of the protocol designed to eliminate differences in method factors and thus heterogeneity---provides at best limited information about---indeed, a lower bound on---heterogeneity. Specifically, because all laboratories use identical materials, any information on heterogeneity provided by this approach relates almost exclusively to differences in the subject populations across the laboratories.

Instead, systematically varying various method factors would provide much more information about the phenomenon in question. For example, when a set of studies features great variation in their method factors and an assessment of heterogeneity is relatively low, this suggests that the phenomenon in question is highly stable with regards to variation in these factors; in other words, these method factors are unlikely to interact with theoretical moderators and the underlying theory may be relatively tight (i.e., in terms of accounting for all relevant theoretical moderators). A large assessment suggests just the opposite, namely that the phenomenon in question is not stable with regards to variation in these factors and the underlying theory may be in need of improvement. On the other hand, when a set of studies features little variation in their method factors and an assessment of heterogeneity is relatively low, this is not particularly informative or probative (although in this setting, a large assessment suggests even more greatly that the phenomenon in question is not stable with regards to variation in these factors and the underlying theory may be in need of improvement).

Consequently, our {\em second recommendation} is that future large scale replication projects systematically vary method factors across the laboratories involved in the project (see also \cite{EhrBou93}, \cite{LinMur93}, and \cite{Barietal18}); ideally, in doing so, they would also take an explicit stance about which ones are likely to be major and which are likely to be minor.

Our {\em third recommendation} is that researchers analyze the data resulting from these projects using a hierarchical model fit to the individual-level observations, and specifically that all theoretical moderators should be modeled  via covariates while all other potential moderators---that is, method factors---should induce variation (i.e., heterogeneity). In classical terms, this would amount to treating theoretical moderators as fixed and method factors as random (though we would not limit ourselves or others to classical models).

Such a modeling approach requires researchers to take an explicit stance on which moderators pertain to theory and which are heterogeneity-inducing method factors (for examples, see \cite{CheBocGoo15} and \cite{McSBoc18}). While one might argue this is difficult or perhaps even unrealistic, we believe it is necessary and beneficial. Carefully-constructed theory involves the delineation of relevant constructs, interactions, boundary conditions, and related matters. The modeling approach we suggest necessitates this thus resulting in clearer and likely improved theory. In addition, the assessments of heterogeneity resulting from this approach have the potential to enrich theory. 

Finally, our {\em fourth recommendation} is that assessments of replicability should not depend solely on estimates of effects, or worse, significance tests based on them. While we do not claim to be able to offer a universal definition of replicability (indeed, we doubt that one single definition would do across domains of research), we believe heterogeneity is an important consideration in assessing replicability. For example, if an effect is estimated to be large but highly heterogeneous, it may be highly replicable according to the classic definition based on sign and statistical significance. However, in our view, sufficiently high heterogeneity does not necessarily indicate replicability of a phenomenon---even if one can be quite sure of its sign. Instead, while recognizing that what constitutes ``sufficiently high'' heterogeneity will vary across domains of research and is a subject matter rather than statistical issue, large variation in effect sizes to us indicates an underlying theory that is poorly constructed and in need of enrichment. In particular, what are believed to be method factors either interact with theoretical moderators or are in fact unaccounted for theoretical moderators. Uncovering this and revising theory in light of it will lead to improved theory that explains heterogeneity and increases replicability.

As an example, consider the anchoring effect, the tendency for individuals to rely too heavily on an initial piece of information offered (known as the ``anchor'') when making decisions. While there is an extremely large literature on this phenomenon (see, for example, Chapters 6-8 of \cite{GilGriKah02}), one popular paradigm for it, used {\em inter alia} in the Many Labs project, is to present subjects with a number that is clearly too small or too large and then to ask them to estimate the distance between San Francisco and New York City; subjects anchor on the number presented and thus tend to provide smaller (larger) estimates when presented with the small (large) number.

Suppose a future large scale replication project follows our recommendations and employs the one phenomenon, many studies approach; systematically varies method factors; analyzes the resulting data using a hierarchical model fit to the individual-level observations with the small versus large anchor specified as the only theoretical moderator; and finds a very large average effect size but large heterogeneity such that the sign of the effect is virtually always in the hypothesized direction. We would not stop here deeming the replication a success.

Instead, we would ask what variation in method factors drove this large heterogeneity. Suppose the researchers examine their data and find that it results from the fact that the studies were conducted across, say, the United States and Europe. Given the operationalization of the dependent measure employed (i.e., estimate of the distance between San Francisco and New York City), it might not be surprising if European subjects relied more heavily on the anchor than American subjects. This may or may not be interesting for theory. On one hand, it could be deemed a mere nuisance interaction between method factors and theoretical moderators---important to know about and perhaps even to account for in the analysis but not of theoretical interest. On the other hand, it could be deemed extremely theoretically interesting: assuming that American subjects are more knowledgeable about the distance between San Francisco and New York City than European subjects, perhaps this result suggests that expertise attenuates or even eliminates the anchoring effect. This could be examined by reformulating the hierarchical model to account for expertise and by designing future studies to explicitly test this.

Of course, if method factors were systematically varied as we recommend, there would be multiple operationalizations of the dependent measure employed (e.g., estimate of the distance between Rome and Berlin). Consequently, one could examine also whether the reverse holds under this operationalization  (i.e., whether American subjects rely more heavily on the anchor than European subjects).

We recognize that conducting multiple replications as opposed to a single replication and systematically varying various method factors is inherently quite challenging in terms of coordination efforts, resources required, and related matters. It also requires more complex designs for replication projects and more elaborate statistical models. Nonetheless, we believe the benefits for assessing replicability (in particular the degree to which method factors moderate the effect in question) and advancing theory are sufficiently valuable to warrant this investment.

\section{Large Scale Replication in Other Domains of Psychological Research}

Because current large scale replications projects have been prospective in nature, they have thus far been restricted to phenomena for which data can be collected quickly and cheaply. For example, one criterion the OSC project authors used to choose the three journals from which they selected phenomena to replicate was that they ``represent psychology sub-disciplines that have a high frequency of studies that are feasible to conduct at relatively low cost'' \cite[]{OSC2015}. Similar criteria applied in the Many Labs project and RRRs. Consequently, these projects have been constrained to certain types of research (e.g., self-report or behavioral data collected from convenience samples) and thus certain domains (e.g., cognitive and social psychology).

Given this, one might ask how to conduct large scale replication projects to assess replicability and understand the role of method factors in other domains of psychological research (e.g., clinical psychology, developmental psychology, behavioral genetics, behavioral neuroscience, health psychology, counseling psychology, and community psychology) where data collection is slow and costly and individual datasets are typically much richer (see also \cite{Tacketal17}). Since researchers in these domains often have access to large amounts of shared (or shareable) archival data, we argue that a retrospective approach to replication is valuable. In particular, researchers in these areas could pool data across laboratories and model it using either a traditional meta-analytic model fit to summary statistics or a hierarchical model fit to the individual-level observations. They could then assess how effects replicate and vary across laboratories and investigate whether method factors either interact with theoretical moderators or are in fact unaccounted for theoretical moderators.

Further, a prospective approach to replication (that is respectful of the pace and cost of data collection in these domains) is also possible and valuable. For example, a researcher planning efforts to investigate some new phenomenon could ensure that variables that speak to previously-investigated phenomena are built into the data collection protocol; in doing so, the researcher could even consciously vary method factors. Then, the new data could be pooled with previously collected data from other laboratories and analyzed via meta-analytic or hierarchical models allowing replicability to be assessed in light of this variation in method factors.

We recognize that barriers to large scale replication in these domains are likely to be quite different from (and larger than) those encountered in current large scale replications projects. For example, a move toward widespread data sharing and thus likely also from single laboratory work to many laboratories work will require shifts in culture, incentives, and infrastructure (e.g., standards for authorship, publication, tenure, and funding). It will also require researchers to better understand statistical methods appropriate for analyzing pooled data (i.e., hierarchical models) and more complex effects (e.g., curve or function estimates as opposed to point estimates); laboratory-specific and other moderators most relevant to include in such analyses; additional method factors that drive heterogeneity (e.g., drop out mechanisms in longitudinal studies); ethical and legal implications of sharing sensitive data; and how to harmonize measurements across laboratories (e.g., if they use different measures of depression). Finally, as discussed in the prior section, assessments of replicability should not depend solely on estimates of effects or significance tests based on them; particularly in the rich datasets common in these domains where there are multiple effects of interest all of which are nonzero but which vary tremendously across laboratories, heterogeneity simply must be a consideration in such assessments.

\section{Discussion}

We have discussed large scale replication projects as a positive development that has emerged from recent difficulties with replication in psychological research. While we believe current large scale replication projects represent an important step forward, we believe they can be made even more informative.

Specifically, replication is complicated in psychological research because studies of a given psychological phenomenon can never be direct or exact replications of one another. As a consequence, heterogeneity cannot be avoided. Rather than trying in vain to eliminate it, we instead argue that large scale replication projects going forward should embrace it through the systematic variation of method factors. 

There is already growing appreciation for the important role that method factors---and the heterogeneity that results from varying them---play in traditional work outside of large scale replication projects. For instance, researchers are increasingly considering how method factors might moderate effects---if only to attempt to explain recent replication failures. More systematically, \cite{SimShoLin17} make the excellent suggestion that all primary research papers include a ``Constraints on Generality'' statement that identifies key method factors, thereby ``help[ing] other researchers...when conducting a direct replication'' and ``encourag[ing] follow-up studies that test the boundary conditions of the original finding.'' Responding to this suggestion, \cite{KenGel18} propose that a Constraints on Generality statement provides the necessary information to allow for inference, via hierarchical modeling and post-stratification \cite[]{ParGelBaf04}, regarding how quantities of interest might vary along with the key method factors identified in the statement. Thus, our recommendations can be seen as part of an increasing call for researchers to think systematically about method factors and heterogeneity.

If our recommendations prove fruitful, perhaps a move to a many phenomena, many studies large scale replication approach may be warranted. This could potentially yield even deeper benefits for theory. Specifically, if seemingly distinct phenomena are subject to the same heterogeneity-inducing method factors, this might point to commonalities at a higher theoretical level.

Our recommendations will not solve recent difficulties with replication in psychological research. Indeed, they are not meant to. Rather, given that heroic coordination efforts and tremendous resources are being invested in large scale replication projects, we would like their payoff to be as large as possible. Large scale replication projects that systematically vary method factors will provide much more information on the degree of heterogeneity and thereby can play an important role not only in assessing replicability but also in advancing theory.

\bibliographystyle{natbib}
\bibliography{lsref}

\end{document}